\documentclass[sigconf, screen, dvipsnames, authorversion]{acmart}

\AtBeginDocument{%
  \providecommand\BibTeX{{%
    \normalfont B\kern-0.5em{\scshape i\kern-0.25em b}\kern-0.8em\TeX}}}

\usepackage{graphicx} 
\graphicspath{{figures/}} 
\usepackage{multirow}
\usepackage{dsfont}
\usepackage{colortbl}
\usepackage[ruled,linesnumbered]{algorithm2e}
\usepackage{booktabs}

\usepackage{float}

\usepackage{caption}
\usepackage{subcaption}

\usepackage{balance}

\usepackage{cleveref}
\crefname{section}{\S}{\S\S}
\crefname{table}{Table}{}
\crefname{figure}{Figure}{}
\crefname{algorithm}{Algorithm}{}
\crefname{equation}{Eq.}{}
\crefname{appendix}{App.}{}
\crefname{prop}{Proposition}{}
\crefname{thm}{Theorem}{}
\crefformat{section}{\S#2#1#3}  

\DeclareMathOperator*{\argmax}{arg\,max}

\newcommand{\indicator}[1]{\mathbb{I}(#1)}

\copyrightyear{2024}
\acmYear{2024}
\setcopyright{acmlicensed}\acmConference[WSDM '24]{Proceedings of the 17th ACM International Conference on Web Search and Data Mining}{March 4--8, 2024}{Merida, Mexico}
\acmBooktitle{Proceedings of the 17th ACM International Conference on Web Search and Data Mining (WSDM '24), March 4--8, 2024, Merida, Mexico}
\acmPrice{15.00}
\acmDOI{10.1145/3616855.3635846}
\acmISBN{979-8-4007-0371-3/24/03}

\begin{document}
\title[Ad-load Balancing via Off-policy Learning in a Content Marketplace]{Ad-load Balancing via Off-policy Learning~\\ in a Content Marketplace}

\author{Hitesh Sagtani}
\affiliation{
\institution{ShareChat}
\country{}
}
\author{ Madan Gopal Jhawar}
\affiliation{
\institution{ShareChat}
\country{}
}
\author{Rishabh Mehrotra}\authornote{Rishabh Mehrotra is now at Sourcegraph.}
\affiliation{
\institution{ShareChat}
\country{}
}
\author{Olivier Jeunen}
\affiliation{
\institution{ShareChat}
\country{}
}

\begin{abstract}

Ad-load balancing is a critical challenge in online advertising systems, particularly in the context of social media platforms, where the goal is to maximize user engagement and revenue while maintaining a satisfactory user experience.
This requires the optimization of conflicting objectives, such as user satisfaction and ads revenue.
Traditional approaches to ad-load balancing rely on static allocation policies, which fail to adapt to changing user preferences and contextual factors.
In this paper, we present an approach that leverages off-policy learning and evaluation from logged bandit feedback.
We start by presenting a motivating analysis of the ad-load balancing problem, highlighting the conflicting objectives between user satisfaction and ads revenue. We emphasize the nuances that arise due to user heterogeneity and the dependence on the user's position within a session. Based on this analysis, we define the problem as determining the optimal ad-load for a particular feed fetch. To tackle this problem, we propose an off-policy learning framework that leverages unbiased estimators such as Inverse Propensity Scoring (IPS) and Doubly Robust (DR) to learn and estimate the policy values using offline collected stochastic data. We present insights from online A/B experiments deployed at scale across over 80 million users generating over 200 million sessions, where we find statistically significant improvements in both user satisfaction metrics and ads revenue for the platform. 

\end{abstract}



\maketitle

\section{Introduction}
Ad-load balancing plays a critical role in content marketplaces, including prominent platforms like Facebook, Instagram, ShareChat and YouTube, where we need to determine the optimal ad-load during a user session.
On one hand, maximizing user satisfaction is crucial to provide a positive user experience and ensure long-term engagement.
On the other hand, advertising revenue is a key factor for the sustainability and profitability of the platform.
The challenge lies in finding the right balance between these objectives. In this paper, we propose a novel approach to ad-load balancing that leverages off-policy learning using unbiased estimators.

In our approach, we define the problem as determining the optimal ad-load for a particular feed fetch, considering the conflicting objectives of user satisfaction and ads revenue.
However, the complexities of the problem go beyond the mere trade-off between these two metrics.
User heterogeneity is an important factor to consider, as different cohorts of users may have varying ad-tolerance levels. Some users may be more accepting of ads, while others may be more sensitive to excessive ad exposure. Furthermore, the impact of other contextual signals adds another layer of complexity. The satisfaction drop caused by increased ad-load may vary depending on \emph{where} in the session the user finds themselves.

Even though this problem is at the heart of online content marketplaces, it has received little attention in the research literature so far.
Indeed, publicly available work does not discuss how advances in machine learning can be leveraged for this common use-case.
To address this gap, we propose an off-policy bandit formulation for ad-load balancing in the context of a social media platform. Our approach leverages user, content, and ads-related features as contextual information within the bandit framework.
We explore different ways to model the action space, going from mere ad \emph{volume} to the \emph{position} of the ads in the feed. 
The contextual bandit paradigm allows us to make informed decisions that balance user satisfaction and advertising revenue, considering user heterogeneity and session context.
We leverage counterfactual optimization techniques on offline collected stochastic data, making use of unbiased estimators based on Inverse Propensity Scoring (IPS) and Doubly Robust (DR) methods~\cite{Saito2022KDD}.
This enables us to evaluate the effectiveness of various ad-load allocation policies in improving both user satisfaction and ads revenue reliably, from offline data.

Finally, we present insights from A/B experiments into the behavior of the selected policies.
We observe directional alignment between our off- and online experiments, in that the top candidates based on offline policy value estimation lead to statistically significant improvements in \emph{both} user satisfaction metrics and advertising revenue compared to homogeneous and static allocation policies, highlighting the value that personalisation can bring.

In summary, the contributions of our work include the following:
\begin{enumerate}
    \item We introduce the ``\emph{ad-load balancing}'' problem, which is at the heart of online content marketplaces but has received little attention in the research literature.
    \item We motivate the problem and its nuances through insights obtained from real-world data (\S\ref{sec:concept}).
    \item We propose an off-policy contextual bandit framework to tackle the problem via counterfactual optimisation (\S\ref{sec:model}).
    \item We present results from live experiments that highlight the effectiveness of our approach, leading to online improvements in both user- and advertiser-focused objectives (\S\ref{sec:applications}).
\end{enumerate}

\begin{figure}
     \centering
     \vspace{-4ex}
     \includegraphics[trim={0cm 0cm 0cm 0cm}, clip, width=0.8\linewidth]{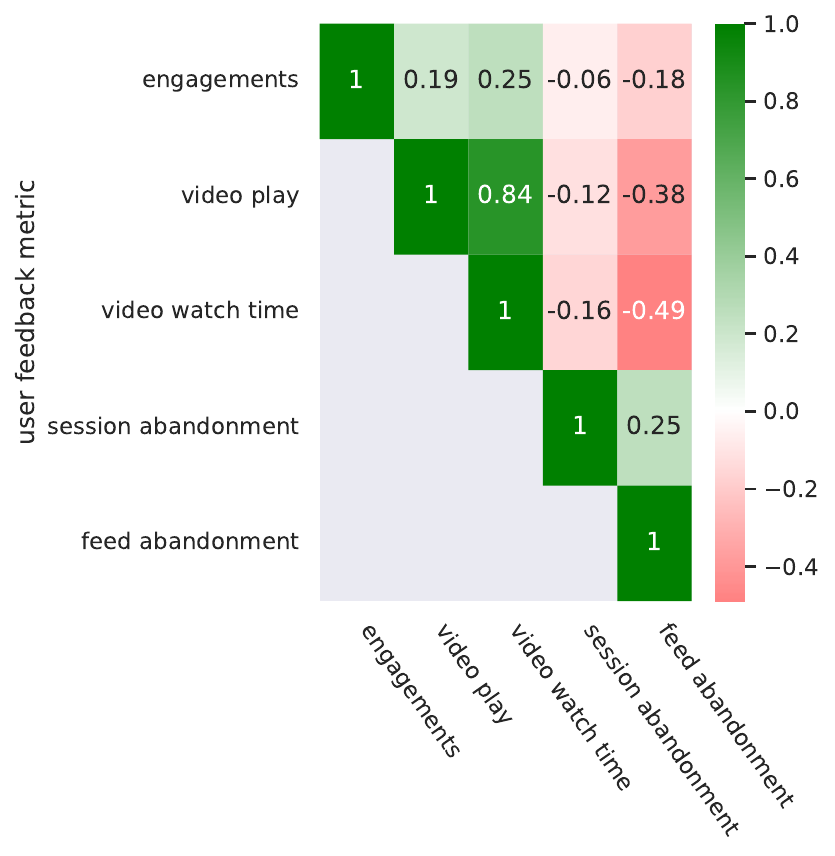}
     \caption{Heatmap visualising Pearson's correlation coefficient among the user-focused objectives we consider.}
     \label{fig:user_sat_correlation}
\end{figure}

\vspace{-2em}
\section{Related Work}

\textit{Ad load personalisation.}
Online advertising is at the heart of the modern web, and a very lively research area as a result~\cite{Bagherjeiran2022}.
The majority of existing work typically focuses on the advertising stack itself, from bidding~\cite{Cai2017,Jeunen2022AuctionGym,Jeunen2023KDD} and response modelling~\cite{McMahan2013,He2014} to auctions~\cite{Liu2021,Jeunen2023AAAI}.
Even though the trade-offs and tensions between advertising load and other platform objectives are widely reported and apparent, works that directly tackle this problem are scarce.
Works exist to tackle ad fatigue from repeated impressions due to the same ad~\cite{Abrams2007}, or general ``banner blindness'' as a result from ad overload~\cite{Resnick2014}.
Other seminal work leverages counterfactual reasoning techniques to gain insights into search engine advertising~\cite{Bottou2013}.
The effects of advertising on user engagement have also been studied in the context of internet radio~\cite{Huang2018}.
To the best of our knowledge, our work is the first that provides a practical and effective method to deal with personalised ad load balancing in online content marketplaces.

\textit{Multi-objective optimisation.}
Advertising problems in content marketplaces inherently deal with multiple stakeholders that each have multiple, possibly conflicting, objectives.
\citeauthor{Abdollahpouri2020} provide an overview of the problems that typically arise in multi-stakeholder recommender systems, and common approaches to solving them~\cite{Abdollahpouri2020}.
\citeauthor{ZHENG2022141} discuss multi-objective optimisation methods proposed in the research literature~\cite{ZHENG2022141}.
\citeauthor{mehrotra2020bandit} propose a bandit-based multi-objective optimisation method geared towards music streaming platforms~\cite{mehrotra2020bandit}, adopting a Generalised Gini Index (GGI) aggregation function to balance multiple objectives~\cite{BusaFekete2017}.
Our work complements this existing research literature by showing how simple scalarisation techniques, which are commonly used in multi-objective contextual bandit use-cases, can be effective in solving the personalised ad load balancing problem.

\textit{Counterfactual learning and evaluation.}
As machine learning use-cases in modern web platforms move from making \emph{predictions} to making \emph{decisions}, a parallel shift is happening from \emph{supervised} learning approaches towards those that leverage \emph{bandit} feedback~\cite{CONSEQUENCES2022}.
Due to the challenges that arise with real-time learning, these systems typically operate in a batch fashion~\cite{Swaminathan2015}.
This line of work is closely related to the offline Reinforcement Learning (RL) literature~\cite{Levine2020}, but practical applications often rely on the \emph{bandit} assumption (i.e. the Markov Decision Process consists of a single timestep or, analogously, there is no \emph{state})~\cite{Bottou2013,Ma2020,Jeunen2022AuctionGym}.
We leverage recent advances in this field to our use-case, and provide a more detailed overview in Section~\ref{sec:OPL}.

\section{Nuances of Ad-load balancing} \label{sec:concept}

In this section, we provide a motivating analysis of the ad-load balancing problem and highlight the challenges it entails: starting with the trade-off between user satisfaction and ads objectives, then discussing heterogeneity among users towards ads tolerance.
To gain an understanding of these challenges, we consider metrics such as retention, user engagement, time spent, ad impressions, and ad clicks to quantify the trade-off in our proposed framework.

\subsection{Data Context}\label{Sec:Dataset}
We collected user data from ShareChat, a widely popular multilingual social media application with over 180 million monthly active users in India, supporting 18 regional languages. Our dataset consists of feed-level data from a randomly selected sample of 5 million users and approximately 250 million feeds.

Typically, a social media platform presents consecutive feed fetches with 10 posts. In order to reduce the action space for off-policy modeling (and hence, the variance of the estimators), we propose to treat a single feed as multiple independent smaller feeds.
This strategy is further elaborated in Section \ref{sec:model-a-ads-selection}. Specifically, we suggest treating a single feed of 10 posts as two independent feeds, each with 5 posts, with the possibility of having 0, 1, or 2 ads.

\subsection{Trade-Off between ads and user satisfaction}\label{sec:trade_off_ads_sat}
To better motivate the need for ad-load balancing, we analyze the global trade-off between ads and user satisfaction metrics. We consider \emph{satisfaction} metrics such as engagement, video play (a binary metric indicating whether a particular video type has been played beyond a specific threshold value), feed depth scrolled (representing the number of successive feed fetches by a user), as well as user \emph{dissatisfaction} metrics like feed abandonment and session abandonment. In terms of advertising, we examine metrics related to ad views and clicks. Figure \ref{fig:user_sat_correlation} shows how such short-term engagement signals are correlated to one another -- and how user satisfaction and dis-satisfaction signals are negatively correlated.

\begin{figure}
     \centering
     \vspace{-4ex}
     \includegraphics[trim={0cm 0cm 0cm 0cm}, clip, width=0.8\linewidth]{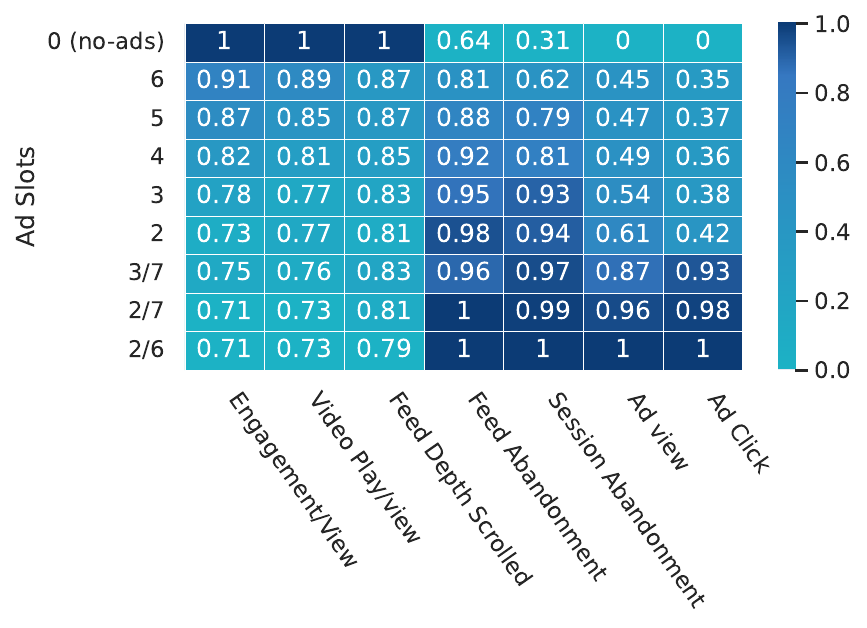}
     \caption{Visualising the inverse relation between user satisfaction \& ads objectives, w.r.t. advertisement positions.}
     \label{fig:user_sat_adslots_variations}
\end{figure}

\textbf{Trade-off with respect to the different ad positions}:  Figure \ref{fig:user_sat_adslots_variations} presents the normalized values of satisfaction, dissatisfaction, and advertising metrics for different ad-slots in a single feed fetch. We observe that increasing the number of ads from 0 to 2 results in decreased satisfaction, increased user dissatisfaction, but also higher ads impressions, clicks, and ultimately (short-term) revenue. Similar effects are observed within ad-slots of 1 ad, when a change in its position from 6 to 2 leads to variations in user metrics due to early exposure to ads at the top of the feed. This phenomenon arises because the view probability decreases for posts further down the feed. To further support these arguments, we present the abandonment analysis in the subsequent section.

\begin{figure}[t]
    \centering
     \vspace{-4ex}
    \includegraphics[width=\linewidth]{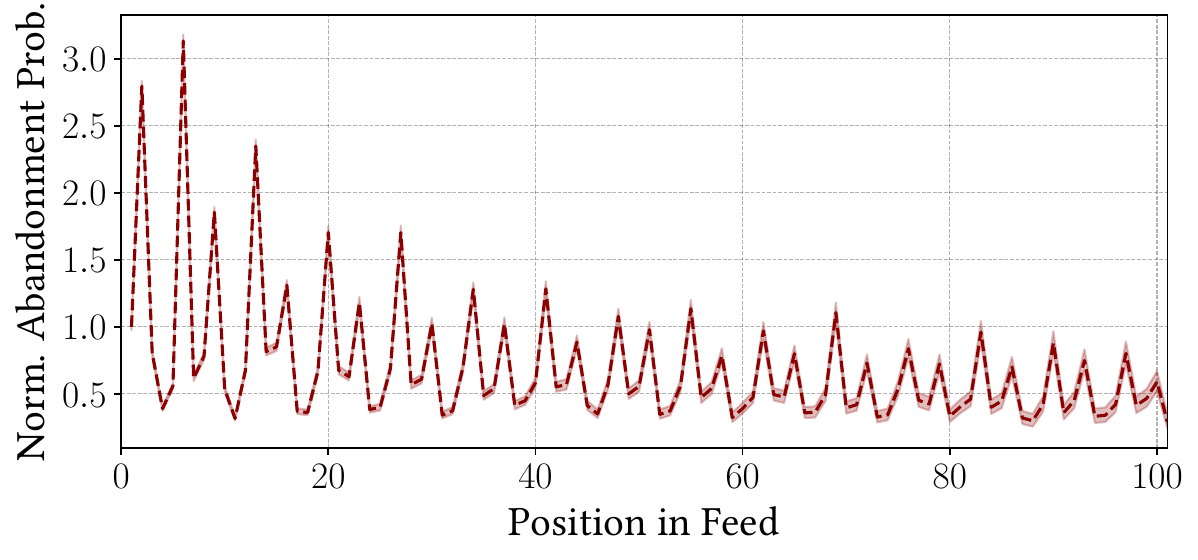}
    \caption{Static advertisement positions lead to increased feed abandonment probabilities (normalised at position 1) right after the ads.}
    \label{fig:abandonment}
\end{figure}

\textbf{Analysing Feed Abandonment}:
A key user dis-satisfaction metric is given by \emph{abandonment} events.
Indeed, if a user leaves the feed, we can interpret this as a negative signal.
Note that this would not always be the case -- as all users, including satisfied ones, leave the feed at some point.
Nevertheless, for the purposes of this analysis we can interpret it as such.
To observe the causal effect that advertisements have on abandonment probabilities, we ran an online test with a static advertising policy and uniformly randomly ranked content.
Indeed, this ensures that the effects of enjoyable content are uniform over the positions, and any observed effects come from other sources (we effectively perform an \emph{intervention}~\cite{Pearl2009}).
As such, if we observe increased abandonment probabilities at the fixed advertisement positions in the feed, this indicates a negative \emph{causal} effect of advertisements on user satisfaction.
Figure~\ref{fig:abandonment} visualises insights from this experiment, with the position on the feed on the x-axis and the abandonment probability on the y-axis. Note that we present \emph{normalized} values of probabilities in the figure (by dividing the actual probability with some constant) to remove commercially sensitivite information. We observe clear negative effects that we would wish to alleviate with more intelligent allocation policies.

\subsection{Heterogeneity across users}\label{sec:fatigue_churn}
\begin{figure}
     \centering
     \vspace{-4ex}
     \begin{subfigure}[b]{0.49\linewidth}
         \centering
         \includegraphics[width=\textwidth]{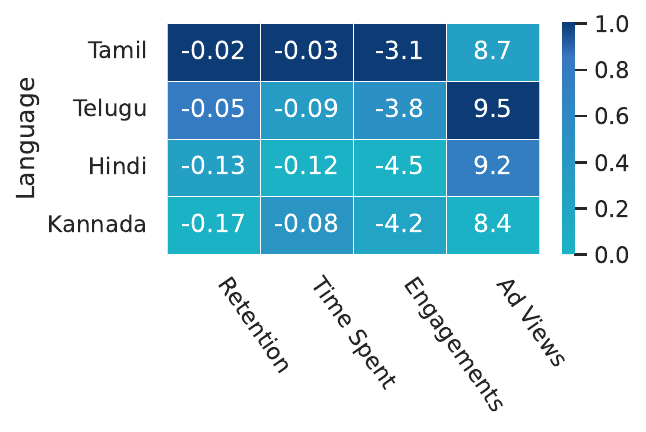}
     \end{subfigure}
     \begin{subfigure}[b]{0.49\linewidth}
         \centering
         \includegraphics[width=\textwidth]{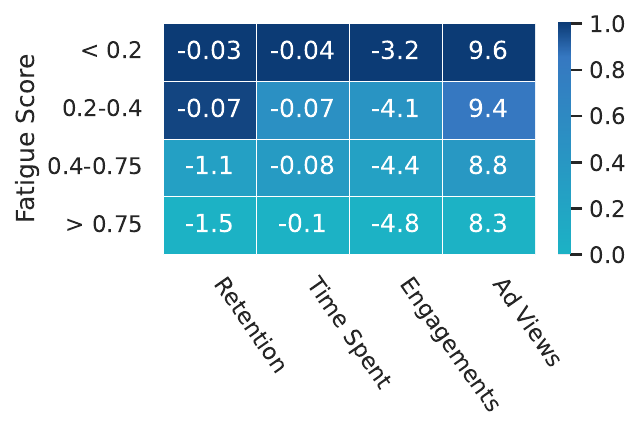}
     \end{subfigure}
        \caption{Visualising user-level heterogeneity to advertising effects, across language and fatigue score.}
        \label{fig:usr_heterogenity_fig}
\end{figure}

In the previous section, we observed the global trade-off between user satisfaction (SAT) and ads metrics. However, it is important to acknowledge that different user cohorts have varying tolerance levels for ad-load, resulting in user heterogeneity. Figure \ref{fig:usr_heterogenity_fig} illustrates the changes in SAT and ads metrics when the ad-load is increased for users in every alternate feed fetch by 1 additional ad. We primarily consider language and fatigue score~\cite{sagtani2023quantifying} to cohort users. The fatigue score was proposed as a surrogate metric for user inactivity on the platform. A lower fatigue score indicates highly active users, while a higher fatigue score suggests inactive users. 

For both user cohorts, we observe an increase in ads impressions after increasing the ad-load in alternate feed fetches. The left section of Figure \ref{fig:usr_heterogenity_fig} demonstrates that the decrease in user satisfaction metrics is less pronounced for Tamil and Telugu users compared to Hindi and Kannada users. Similarly, for users with lower fatigue scores, the decrease in user satisfaction metrics is less significant compared to users with higher fatigue scores. Additionally, we note that ads impressions \emph{decrease} as fatigue score increases, as more active users tend to perform more feed scrolling, resulting in increased impression opportunities. These findings highlight the heterogeneity in ad-tolerance among user cohorts and emphasize the importance of incorporating user context to determine the appropriate ad-load.

\begin{table*}[t!]
\centering
 \vspace{-3ex}
\caption{Optimally learned weights for user (dis-)satisfaction metrics and advertising objectives in our reward function.}
{
\begin{tabular}{ lll}
\specialrule{.1em}{.05em}{.05em}
\textbf{Signal} & \textbf{Weight} & \textbf{Description} \\ 
\specialrule{.1em}{.05em}{.05em}

\textit{\underline{User (dis-)satisfaction signals}} & & \\

Engagements & 0.5995 & User's engagement signals such as likes, sharing on other platforms \& downloads \\
Video play & 0.6235 & A binary metric indicating whether a particular video type has been played \\
Percentage video watch & 0.3464 & A continuous variable indicating the fraction of the video watched by the user \\
Feed depth scrolled & 0.3213 & The number of successive feed fetches by a user \\
Video skip & -0.1432 & A binary label, positive if a video watch is less than 2 seconds  \\
Discounted feed abandonment & -0.3742 & Did the user quit the feed, but enter another feed after the ad? \\
Discounted Session abandonment & -1.2345 & Did the user quit the session entirely after the ad? \\
\hline

\textit{\underline{Ads objective signals}} & & \\
Impression & 0.2234 & Did the user see the ad in case of CPM campaign? \\
Clicks & 0.5135 & Did the user click on the ad in case of CPC campaign? \\
Install & 0.7823 & Did the user install the app in case of CPI campaign? \\
\hline

\end{tabular}
}
\label{tab:reward_weights}
\end{table*}

\vspace{-1em}

\section{Problem Formulation} \label{sec:model}
Contextual and personalised ad-load balancing is an important and common problem -- but data-driven approaches have not been reported in the research literature.
As we effectively aim to maximise cumulative \emph{rewards} by assigning the right \emph{actions} to \emph{contexts}, the problem setting matches the contextual bandit paradigm.
On-policy bandit approaches that learn online are ill-suited for our setting, as they might show unpredictable behaviour once deployed~\cite{akker2023practical}.
Instead, the logged data mentioned in Section \ref{Sec:Dataset} allows us to leverage advances in off-policy learning for our use-case.
We leverage three well-known families of methods: the Direct Method (DM), Inverse-Propensity-Weighted (IPW) empirical risk minimisation (ERM), and a doubly robust variant \cite{dudik2011doubly, su2019cab, swaminathan2015batch, swaminathan2015self}.
In what follows, we introduce these methods, and explore various options when designing the action space and the reward function.

\subsection{Propensity score validation}
We collect data from a uniform random logging policy in the form (\textit{x, a, r$_a$, p$_a$}), where $x \sim D(x)$ is the observed context, $a \in A$ is the action drawn from a uniform random distribution over the action space, r$_a$ is the corresponding reward we observe and p$_a$ is the propensity score of selecting an action.

Off-policy evaluation and learning approaches that rely on importance sampling or IPW require \emph{full support} of the logging policy~\cite{Owen2013}.
This implies that the probability with which the logging policy selects actions should be non-zero for every possible context-action pair: $ p_{a} > 0 \ \forall a \in A $.
This requirement is easily verified for the uniform random policy, as $ p_{a} = 1/ \left | A \right | $ where \textit{A} is the action space.
Aside from this explicit assumption -- practical applications often prohibit effective unbiased reward estimation due to a variety of reasons~\cite{London2022,akker2023practical}.
\citeauthor{mehrotra2020bandit}~\cite{mehrotra2020bandit} and \citeauthor{li2015counterfactual}~\cite{li2015counterfactual} propose two simple tests that validate the logged propensity scores, which we explicitly adopt here.

\textbf{Arithmetic Mean Test:} To assess the accuracy of the randomized data collection process, we examine the frequency of a specific action $a$ in the data and compare it to the expected number of occurrences based on the logged propensity scores. Our analysis reveals that the observed difference is not statistically significant, suggesting that there are no errors in the randomized data collection process.

\textbf{Harmonic Mean Test:} We verify the mean of the random variable in Equation~\ref{eq:harmonic_mean} below is close to 2.
\begin{equation}\label{eq:harmonic_mean}
\mathbb{E}_a \left[ \dfrac{\indicator{a=a^*}}{p_{a^*} } + \dfrac{\indicator{a\neq a^*}}{1- p_{a^*} } \right] \approx 2
\end{equation}

Note that the control variate test proposed by \citeauthor{London2022} is not directly applicable in our setting as it requires a fixed target policy, which we are yet to learn~\cite{London2022}.
Additionally, note that this type of logging policy is devoid of either observed or unobserved confounding by design~\cite{Jeunen2023_confounding}.

\begin{table*}[t!]
\centering
 \vspace{-3ex}
\caption{Type and description of some of the most important features used to represent contextual signals in our bandit setup.}
{
\begin{tabular}{ lll}
\specialrule{.1em}{.05em}{.05em}
\textbf{Attribute Type} & \textbf{Name} & \textbf{Description} \\ 
\specialrule{.1em}{.05em}{.05em}

\multirow{5}{*}{User (Dynamic)} & Current Interactions & Counters of hourly and daily interactions like engagement, time spent, and views.  \\ 
& Activity & Login counters like average inactivity last week, logins yesterday, etc. \\ 
& Historical Interactions & Counters over the last 3, 5 and 7 day window. \\ 
& Embeddings & Dot product between pre-computed user embedding and average of post embedding in the feed. \\ 
& Fatigue Score & Described in section \ref{sec:fatigue_churn} -- proposed by~\cite{sagtani2023quantifying} \\ 
\hline

\multirow{3}{*}{User (Static)}  & Platform Age & Number of days since user signed up on the platform. \\ 
& Language & One-hot encoding of the user language. \\ 
& Location & One-hot encoding of state, district, and city. \\ 
\hline

\multirow{3}{*}{Content} & Genre & A genre affinity score and the number of posts in the feed. \\ 
& Distinct Genres & Taken as proxy for diversity, as the number of distinct genres in the feed. \\ 
& Post Age & Average age of posts in the feed. \\ 
\hline

\multirow{6}{*}{Advertisements}  
& Previous Ad Slots & Ad-slots in the previous feed request. \\ 
& Ad Gap & Number of posts between the last ad in the previous feed and the first ad in the current feed. \\ 
& Average Ad Load & average number of ads per feed in last 3, 5 feed fetches \\
& Total Ad Impressions & Number of ad impressions for this user in the current session. \\
& IAB Category & Interactive Advertising Bureau (IAB) content taxonomy relating to the ad.\\  
& Clicks \& Impressions & The user advertising impressions and clicks in several time windows. \\ 
\hline

\end{tabular}
}
\label{tab:classification_features}
\end{table*}

\subsection{Off-policy bandit formulation}\label{sec:model-a}
\subsubsection{ \textbf{Designing the action space}}\label{sec:model-a-ads-selection}
Two natural choices arise when considering \emph{actions} in the ad-load balancing problem:

\textbf{Volume of advertisements}.
We can consider fixed ad positions in the feed, and model the number of ads we wish to show in a given feed fetch: $0, 1, 2$.
A clear advantage of this approach is its simplicity -- but it lacks the expressiveness to explicitly tackle the position effects we observe in Figure~\ref{fig:user_sat_adslots_variations}.

\textbf{Volume and position of advertisements}.
To overcome the limitation of the first approach, we model an action as \emph{both} the number of ads as well as the position of the ads in the feed fetch.
This accounts for the heterogeneous distribution in the expected rewards for different ad positions.
One challenge with this approach is the combinatorial explosion of the size of the action space.
In general, for a feed of length \textit{n} and fixed number of \textit{i} ads in the feed for $i \leq n$, the total number of arms is given by the binomial coefficient ${n\choose i}$.
As such, the total number of possible combinations for all possible ad loads and positions are given by $\sum_{i=0}^{n} {n\choose i} = 2^n$.
With exponentially many actions, the propensity of each action becomes $1/2^n$ when following a uniform logging policy.
Although the reward estimates from IPW estimators are still unbiased, their variance quickly becomes problematic.
To deal with this effectively, we include two further considerations:

\begin{enumerate}
    \item The standard length of a feed fetch $n$ is $10$, which would imply an action space of $1024$. We treat a feed of 10 length as two independent feeds of $5$ length each, allowing us to reduce the size of the action space by a factor $2^5=32$ to $32$. 
    \item We analyzed the effect of increasing the gap between consecutive ads on user satisfaction metrics. When the gap between consecutive ads was $\leq 3$ or an ad was placed at $1^{st}$ position, we observed significant satisfaction losses, which are undesirable for the platform. Therefore, we removed such combinations from the action space.    
\end{enumerate}

\subsubsection{\textbf{Designing the reward function}}
The ad-load balancing problem is inherently multi-stakeholder and multi-objective.
Indeed, \emph{users} and \emph{advertisers} have their own objectives which can be conflicting.
We define an explicit reward metric for both:

\textbf{User Satisfaction} is not straightforward to measure from implicit feedback.
The core hypothesis many platforms make, is that \emph{retention} reflects users' satisfaction with the system.
We model this with a binary label: ``\emph{do users come back to the system tomorrow?}'' and call it \emph{D1 Retention}.
Because retention is a long-term and delayed signal, adopting it directly as a reward can be challenging.
As such, we consider feed-level user (dis-)satisfaction signals mentioned in table \ref{tab:reward_weights} to model a reward that serves as a proxy metric for retention. Let $\delta_1$ be user's \emph{D1 Retention}, $\phi{(x, a)}$ be the user SAT metric, and $x_i$ be the value of $i^{th}$ (dis-)satisfaction signal with weight $w_i$. We adopt linear scalarization to estimate \textit{D1 retention} from these different signals using Equation~\ref{eq:scalarisation}, where $\rho$ indicates Pearson's correlation: 

\begin{equation}\label{eq:scalarisation}
\begin{split}
    \phi^{\star} = \argmax_{\phi} \rho(\delta_1, \phi{(x, a)}), \text{ with } \phi{(x, a)} = \sum_{i} w_i \cdot x_i.
\end{split}
\end{equation}

The objective in Equation~\ref{eq:scalarisation} is maximised by learning a linear model via gradient ascent on empirical logged data. The optimal weights obtained for each signal $x_i$ in the user SAT metric $\phi{(x, a)}$ are provided in Table \ref{tab:reward_weights}, along with corresponding signal descriptions. Notably, positive signals hold positive weights, while negative signals carry negative weights in the final user SAT reward. Finally, we also emphasize certain considerations to keep in mind while designing the reward for abandonment signals below.

\textit{Discounting \& Attribution of Abandonment Dissatisfaction Signal}: Usually, users scroll through feeds continuously until they either switch to a different feed or exit the session. We define $rank_{i}$ as the user's current feed fetch count in a series of consecutive feed fetches and $rank_{d}$ is the total number feeds scrolled consecutively where $rank_{i}<=rank_{d}$, then user abandons the feed at $rank_{d}$. As users leave a sequence of feeds at varying points, dissatisfaction from leaving the first feed is more pronounced than leaving, say, the $10^{th}$ feed. This is because users have already consumed around 50 posts by the $10^{th}$ feed. So, the abandonment cost should be lower at higher feed depths. Additionally, some of the abandonment cost at a particular feed depth may or may not be attributed to the previous feed depths. If $\lambda$ is the final feed abandonment signal, we account for both discounting and attribution as:
\begin{align}\label{eq:discounted_feed_abandonment}
    \lambda &= \text{Discounted Signal} \cdot \text{Signal Attribution} \\ 
    &= \frac{1}{\log(1+rank_{d})} * \alpha^{rank_{d} - rank_{i}}
\end{align}
where 0 < $\alpha$ <= 1, is a hyper-parameter controlling the degree of attribution. If $\alpha=1$, full cost is attributed to the previous feeds, and if $\alpha  \rightarrow 0^{+}$, almost no cost is attributed to the previous feed fetches.

Similar to feed abandonment, we apply a discount to the session abandonment signal depending on the amount of time the user has already spent in the session.

\textbf{Advertising objectives} can be manifold but are typically focused on revenue.
Nevertheless, similar to true user satisfaction, advertising revenue is a long-term metric that is not easily plugged in directly as a reward.
As short-term proxies, we adopt a scalarised version of advertising impressions, clicks and installs that maximises the pearson correlation with revenue. Let $Rev$ be \emph{revenue}, $\psi{(x, a)}$ be the ads-objective, and $x_i$ be the value of $i^{th}$ ads objective signal with weight $w_i$. We adopt linear scalarization to estimate \textit{revenue} from these different signals using Equation~\ref{eq:ads_scalarisation}, where $\rho$ indicates Pearson's correlation coefficient:

\begin{equation}\label{eq:ads_scalarisation}
\begin{split}
    \psi^{\star} = \argmax_{\psi} \rho(\text{Rev}, \psi{(x, a)}), \text{ with } \psi{(x, a)} = \sum w_i * x_i.
\end{split}
\end{equation}

Similar to user SAT optimization, Equation~\ref{eq:ads_scalarisation} is maximised by learning a linear model via gradient ascent on empirical logged data. The optimal weights obtained for each signal $x_i$ in the ads objective $\psi{(x, a)}$ are provided in Table \ref{tab:reward_weights}, along with corresponding signal descriptions.

\textbf{Final reward.} Let $\phi^{\star}{(x, a)}$ and $\psi^{\star}{(x, a)}$ be the optimal user SAT metric and advertising objective respectively obtained from methods above, then the final reward is a weighted combination of both as follows:
\begin{equation}\label{eq:final_reward_eq}
\begin{split}
    R = \beta*\phi^{\star}{(x, a)} + (1-\beta)*\psi^{\star}{(x, a)},\\ 
    \text{with } \beta \in [0, 1].
\end{split}
\end{equation}

\subsubsection{\textbf{Designing a context representation}} 
\begin{figure}
     \centering
     \vspace{-2ex}
     \includegraphics[width=0.8\linewidth]{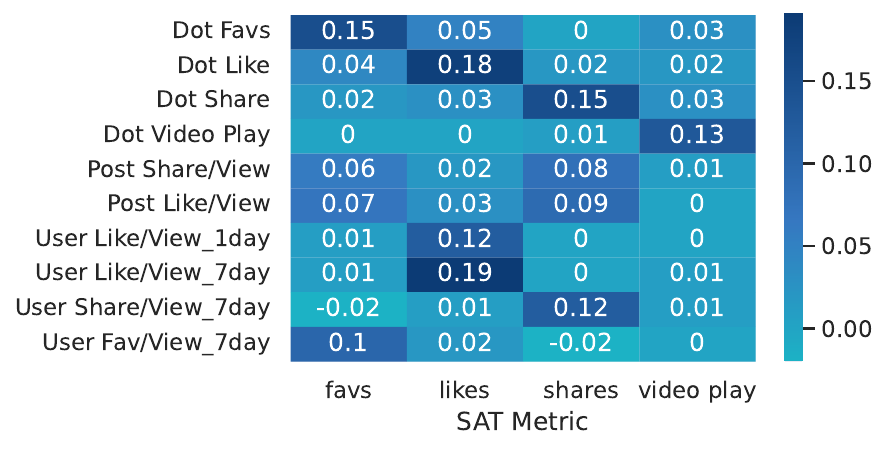}
     \caption{Heatmap of Pearsons' correlation coefficients between features and satisfaction signals.}
     \label{fig:corrs_fts_sat}
\end{figure}

We use user, content and advertisement attributes as contextual signals. Some of the features along with the description are listed in Table \ref{tab:classification_features}. As part of the feature pre-processing we removed the features with high multicollinearity.
Figure \ref{fig:corrs_fts_sat} visualises a heatmap of correlation of important features with our user SAT reward signals. Note, \textit{Like}, \textit{Shares} \& \textit{favs} (download) are the different types of engagement signal users can do with a post. Dot features highlighted in the correlation heatmap are the dot product between the pre-trained embedding of users and post on a particular signal and per view attributes are the counter attributes for a user/post in a specific time window. For eg: \textit{$User Like/View\_1day$} mean the total number of likes / total number of views of a user on the platform in 1 day window.
Their high correlation with individual reward signals for engagements (e.g. 0.18 for ``likes'') underline the utility of such features in our dataset.

\begin{figure}
     \centering
     \vspace{-2ex}
     \includegraphics[width=0.7\linewidth]{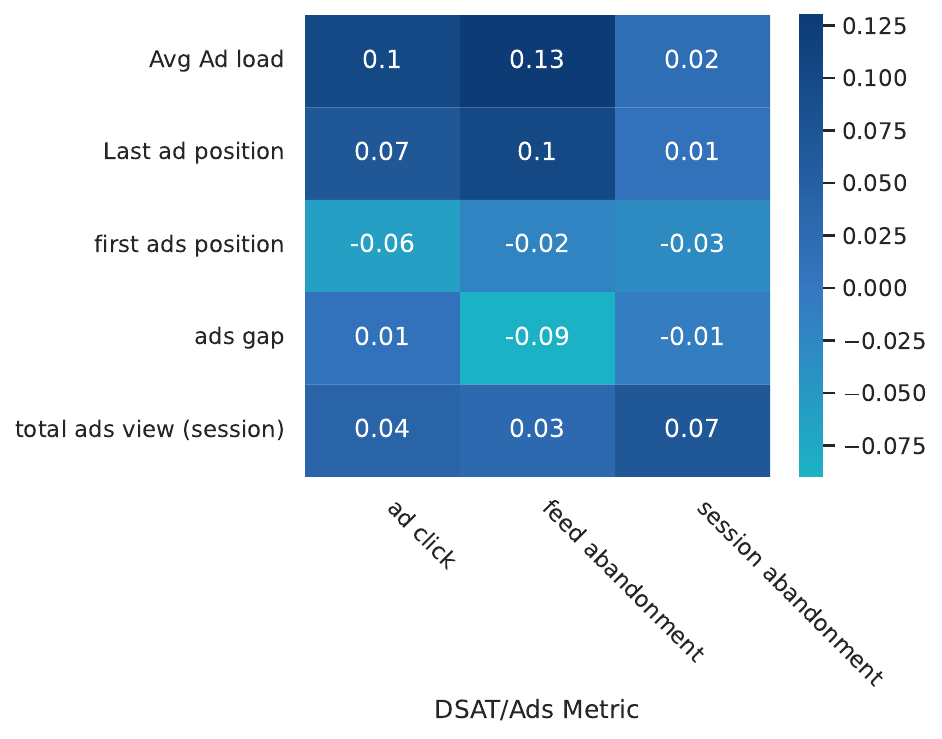}
     \caption{Heatmap of Pearson's correlation coefficients between features and dissatisfaction / advertising signals.}
     \label{fig:corrs_fts_ads}
\end{figure}

Figure \ref{fig:corrs_fts_ads} illustrates the correlation between advertisement features and both the advertising objective and the user dis-satisfaction signal. Alongside traditional counter features (impressions/clicks on an advertisement in various time windows) and IAB category features, real-time features such as ad-slots displayed in the last feed fetch and the average ad-load in the current session are important and their significance is highlighted by their strong correlation with clicks.
A high positive correlation between the number of ads, representing an ad-heavy feed, leads to higher feed- and session-abandonment, but also more advertising clicks.
This makes the trade-off inherent to our problem apparent.

\subsection{Off-policy learning and evaluation}\label{sec:OPL}
The research literature has conventionally focused on so-called \emph{on-policy} bandit approaches, where a policy is deployed and allowed to learn and update in real-time.
Although the advantages of this approach are apparent --- it should be clear that it brings significant challenges.
Indeed, not only does it require significant engineering bandwidth to set up the proper infrastructure, we essentially would have no way of properly vetting the ad load balancing policy before deployment.
For these reasons, \emph{off-policy} bandits are generally preferred in real-world applications~\cite{Chen2019, Ma2020}.

Off-policy learning is often called \emph{counterfactual} learning -- as it aims to optimise a policy for a counterfactual estimate of the reward that policy would have collected.
A \emph{policy} is a contextual probability distribution, often obtain through a parametric model.
We will denote such a parametric policy with shorthand notation $\pi_{\theta}(a|x) \equiv \mathsf{P}(A=a|X=x;\Pi=\pi_{\theta})$.

We wish to learn the parameters $\theta^{\star}$ that maximise the estimated expected value we obtain under the policy $\pi_{\theta^{\star}}$ on logged data $\mathcal{D}$:
\begin{equation}
    \theta^{\star} = \argmax_{\theta \in \Theta} V(\pi_{\theta}, \mathcal{D}).   
\end{equation}

Several families of estimators exist for $V$.

\paragraph{The Direct Method (DM)}
So-called \emph{value}-based methods leverage supervised learning methods for reward estimation.\footnote{These are often referred to as Q-learning in the reinforcement learning literature.}
In the direct method, the parameters $\theta$ are used to learn a model for the reward an action yields, given a context: $\widehat{r}(a|x) \approx \mathbb{E}[R|X = x, A = a]$.
The reward model $\widehat{r}(a|x)$ can then be used to estimate the policy value:
\begin{equation}\label{eq:dm}
    \widehat{V}_{\text{DM}}(\pi_{\theta},\mathcal{D}) = \sum\limits_{(x, a, r_{a}, p_{a}) \in \mathcal{D}} \sum\limits_{a^{\prime} \in A} \pi_{\theta}(a^{\prime}|x) \cdot \widehat{r}(a^{\prime}|x).
\end{equation}
From Eq.~\ref{eq:dm}, we can observe that the optimal policy for a given reward regressor $\widehat{r}_{\theta}(a|x)$ is the deterministic policy that chooses the action with the highest reward estimate: $ \argmax_{a^{\prime}\in A} \widehat{r}(a^{\prime}|c)$.
As this foregoes the need of sampling from a policy, they are widely used in practice.
Then, the parameters ${\theta}$ are used for $\widehat{r}$, as $\pi$ reduces to the simple decision rule laid out above.
Although value-based methods generally have low variance, they are biased estimators.
Advances exist in the research literature to improve on such methods, typically by adopting some form of \emph{pessimism} in the reward model~\cite{Mykhaylov2019CausalML,Jeunen2021A,Jeunen2023TORS}.

\paragraph{Inverse Propensity Weighting (IPW)}
\emph{Policy}-based methods directly learn a parametric policy, foregoing the need to learn an explicit reward model.
Inverse Propensity or Weighting (IPW) is the workhorse behind this line of work~\cite{Bottou2013}.
Using the logged propensities $p_{a}$, we can obtain an unbiased estimator of the reward that a new policy $\pi$ would have obtained on a logged dataset $\mathcal{D}$:
\begin{equation}\label{eq:ips}
    \widehat{V}_{\text{IPW}}(\pi_{\theta},\mathcal{D}) = \sum\limits_{(x, a, r_{a}, p_{a}) \in \mathcal{D}} r_{a} \cdot \frac{\pi_{\theta}(a|x)}{p_{a}}.
\end{equation}
Although it is easy to see that Equation~\ref{eq:ips} is unbiased, it typically leads to excessive variance.
Extensions of this estimator typically aim to handle that directly: by capping~\cite{Ionides2008} or self-normalising~\cite{Swaminathan2015snips,Joachims2018} the weights, or adding regularisation terms to the objective~\cite{Maurer2009,Swaminathan2015,Ma2019AIStats,Faury2020,Si2020,JeunenKDD2020}.
Similarly to the value-based family, these extensions can be interpreted as mathematical \emph{pessimism}~\cite{Jeunen2023TORS}.
Recent work has shown that \emph{gradient boosting} techniques and models (such as Gradient Boosted Decision Trees (GBDTs)~\cite{Prokhorenkova2018}) are amenable to such policy learning objectives~\cite{London2023}.

\paragraph{Doubly Robust (DR)}
The value- and policy-based families both have their own advantages and characteristics.
\citeauthor{dudik2011doubly} introduce an estimator that leverages \emph{both} a reward model $\widehat{r}$ \emph{and} the logging propensities $p_{a}$: the Doubly Robust estimator~\cite{dudik2011doubly}.
It derives its name from a desirable property, in that it is unbiased if \emph{either} the logged propensities, or the reward model are.
\begin{equation}\label{eq:doublyrobust}
\begin{gathered}    
    \widehat{V}_{\text{DR}}(\pi_{\theta},\mathcal{D}) = \\
    \sum\limits_{(x, a, r_{a}, p_{a}) \in \mathcal{D}} \left( (r_{a}-\widehat{r}(a|x)) \cdot \frac{\pi_{\theta}(a|x)}{p_a} + \sum\limits_{a^{\prime} \in A} \pi_{\theta}(a^{\prime}|x) \cdot \widehat{r}(a^{\prime}|x) \right)
\end{gathered}
\end{equation}
DR extends DM with an IPW-term that weights the error of the reward regressor.
Extensions exist that learn $\widehat{r}$ to minimise DR's variance~\cite{Farajtabar2018}, optimise the DM--IPS balance~\cite{su2019cab}, or introduce further transformations that bound expected error~\cite{Su2020_ICML}.
Other work has shown that DR can still be outperformed by either IPW or standalone DM~\cite{Jeunen2020REVEAL}.
In other words, there is ``\emph{no free lunch}''.

\section{Experimentation} \label{sec:applications}
In order to empirically validate the effectiveness of our proposed framework through experimentation, we would require stochastic logged data with user and advertising context, advertising positions with associated propensity values, as well as user feedback on both ads and non-ads content.
To the best of our knowledge and at the time of writing, no such datasets are publicly available.
Hence, we need to resort to proprietary datasets.
We note that our proposed framework is general, and makes few assumptions about the nature of the problem setting.
As a result, we believe that our introduced framework and results can extend to other platforms with similar or different feed models~\cite{chuklin2015click, Jeunen2023_C3PO}.

\subsection{Policy Design}
To achieve a balanced trade-off between user satisfaction and ads objectives, we propose various policy designs, including both baseline and counterfactual-based approaches.

\subsubsection{\textbf{Baseline Policies}}\label{sec:baselines}

We begin by defining a range of baseline policies, including heuristic-based and learned strategies to jointly optimize user satisfaction (SAT) and ads objectives.

\begin{enumerate}
    \item \textbf{Optimizing User SAT}: The first policy we investigate optimises only user satisfaction by setting $\beta=1$ in equation \ref{eq:final_reward_eq}. Optimizing such a policy leads to no ads for every feed fetch, aligning with offline observations where ad-free slots yield highest user SAT metrics (Figure \ref{fig:user_sat_adslots_variations}).

    \item \textbf{Optimizing Ads Objective}: We solely optimize the ads objective by setting $\beta=0$ in equation \ref{eq:final_reward_eq}. While optimizing for the user SAT gives no ads at all, this approach maximize the ads with minimal user SAT. Our experiments consistently yield an ads position of ``2 and 6'', in line with offline ad reward patterns (Figure \ref{fig:user_sat_adslots_variations}).

    \item \textbf{Random Policy}: We move from single-objective optimization to policies considering both user SAT and ads objectives. We start with a random policy which selects an arm from the action space using a uniform random distribution.

    \item \textbf{Static Policy}: We define static policy as a function of offset and $post\_gap$ (post\_gap is the number of non-ad posts between consecutive ads), wherein offset is the position of the ad on the very first feed fetch by the users after which ads are displayed with a fixed $post\_gap$. For instance, a static policy of $(3, 5)$ positions the first ad at ``3'', maintaining a consistent gap of 5 non-ad posts.

    \item \textbf{User Fatigue Based Policy}: The baselines we have looked so far are either hard coded heuristics w.r.t. advertising \emph{load} and \emph{positions}, or they optimize for single objective. We leverage a method based on the fatigue score (learned using GBDT model) ~\cite{sagtani2023quantifying} to optimize both the user SAT and ads-objective. Let $\phi_u$ represents the user's fatigue score, $\theta$ represents the default number of ads shown to users in a feed on the platform, and $\theta_u$ represents the updated number of ads shown to the user. Then, the fatigue score-based policy can be formulated as:
    \[
        \theta_u = \begin{cases}
        \theta + 1, & \text{if } \phi_u < \alpha\\
        \theta, & \text{if } \alpha < \phi_u < \beta\\
        \theta - 1, & \text{if } \phi_u > \beta \\
    \end{cases}
    \]
    where $0 < \alpha < \beta < 1$ are the thresholds for low and high fatigue users because of the ads respectively. Users with fatigue scores because of ads below $\alpha$ experience an increased ad load, while scores above $\beta$ result in a reduced ad load. 
    As this baseline optimises both user SAT and advertising objectives in a personalised manner, it is a much stronger baseline than (1)--(4). This is reflected in our empirical results.
\end{enumerate}

\subsubsection{\textbf{Policies based on counterfactual estimators}} 
In addition to the aforementioned baselines, we optimize policies parameterised by Multi-Layer Perceptrons (MLPs) and Gradient-Boosted Decision Trees (GBDTs) using the unbiased estimators (IPW and DR, see Section~\ref{sec:OPL}) from logged data. We fine-tune the reward function from equation \ref{eq:final_reward_eq} across various $\beta$ values. Additionally, alongside propensity-based methods, we train MLPs and GBDTs to predict the reward function, known as the Direct Method.

\begin{table*}[t!]
\centering
 \vspace{-3ex}
\caption{Online A/B: \% change in user SAT and ads metrics w.r.t. control. All results are
statistically significant (2 tailed t-test at p<0.05 after Bonferroni correction) except for those marked by$^+$.}
{
\begin{tabular}{ ccccccccccc}
\specialrule{.1em}{.05em}{.05em}
&  & & \multicolumn{5}{c|}{SAT Metrics} & \multicolumn{3}{c}{Ads Metrics} \\
\cline{4-11}
Variant & $\beta$ & Objective & D1 Retention & Time spent & Views & Engagements & Video play & Impressions & Clicks & Revenue \\
\specialrule{.1em}{.05em}{.05em}

variant-1 & 0.7 & \textit{DR} & -0.01$^+$ & -0.01$^+$ & -0.1 & -0.24 & -0.07$^+$ & 1.14 & 1.91 & 1.65 \\
variant-2 & 0.8 & \textit{DR} & 0.04$^+$ & 0.14 & 0.02$^+$ & -0.09$^+$ & 0.19 & 0.79 & 2.05 & 1.45 \\
variant-3 & 0.9 & \textit{DR} & 0.08 & 0.22 & 0.15 & -0.02$^+$ & 0.31 & 0.28 & 0.52 & 0.2 \\
\hline

\end{tabular}
}
\label{tab:ab_online_results}
\end{table*}

\subsection{Offline Experiments}
The Open Bandit Pipeline library \cite{saito2020open} was used for training and evaluating these policies.
For hyperparameter tuning of the supervised models, we performed a randomized grid search over various combinations and selected the parameters that yielded the best performance on the validation set.
We report the optimal hyperparameters for our GBDT and MLP models in the supplementary material.

Figure \ref{fig:offpolicy_eval} illustrates the percentage loss in Satisfaction (SAT) and ads objectives for a particular policy compared to the optimal policy described under section \ref{sec:baselines} for each objective in isolation. The losses are calibrated on a scale of 0-100 for ads and 100-0 for no ads. The \emph{offset} of the static policy are highlighted in the figure and have a ``post gap''$=5$. From the plot, we observe that the baseline policies exhibit higher SAT and ads loss compared to the counterfactual policies, indicating their Pareto inefficiency. 
Among the baselines, the fatigue score-based policy performs best, exhibiting the lowest loss in both SAT and ads reward. This result highlights the potential of personalization in attaining better outcomes over static policies while simultaneously optimizing both objectives.

Among the counterfactual based policies, by varying the $\beta$ value (from equation \ref{eq:final_reward_eq}) from $0.7$ to $0.9$ we observe a decrease in user SAT loss and an increase in ads objective loss. This suggests that $\beta$ serves as a trade-off parameter between SAT and ads objectives, with all points lying on the same Pareto front. For $\beta=0.8$, both GBDT and MLP models exhibit similar losses in both SAT and ads objectives. Additionally, we observed that policies trained with the DM objective were Pareto inefficient compared to policies trained with the IPW and DR objectives. The convergence of IPW and DR indicates the reliability of our propensities to estimate rewards.

\begin{figure}[t]
    \centering
     \vspace{-2ex}
    \includegraphics[width=\linewidth]{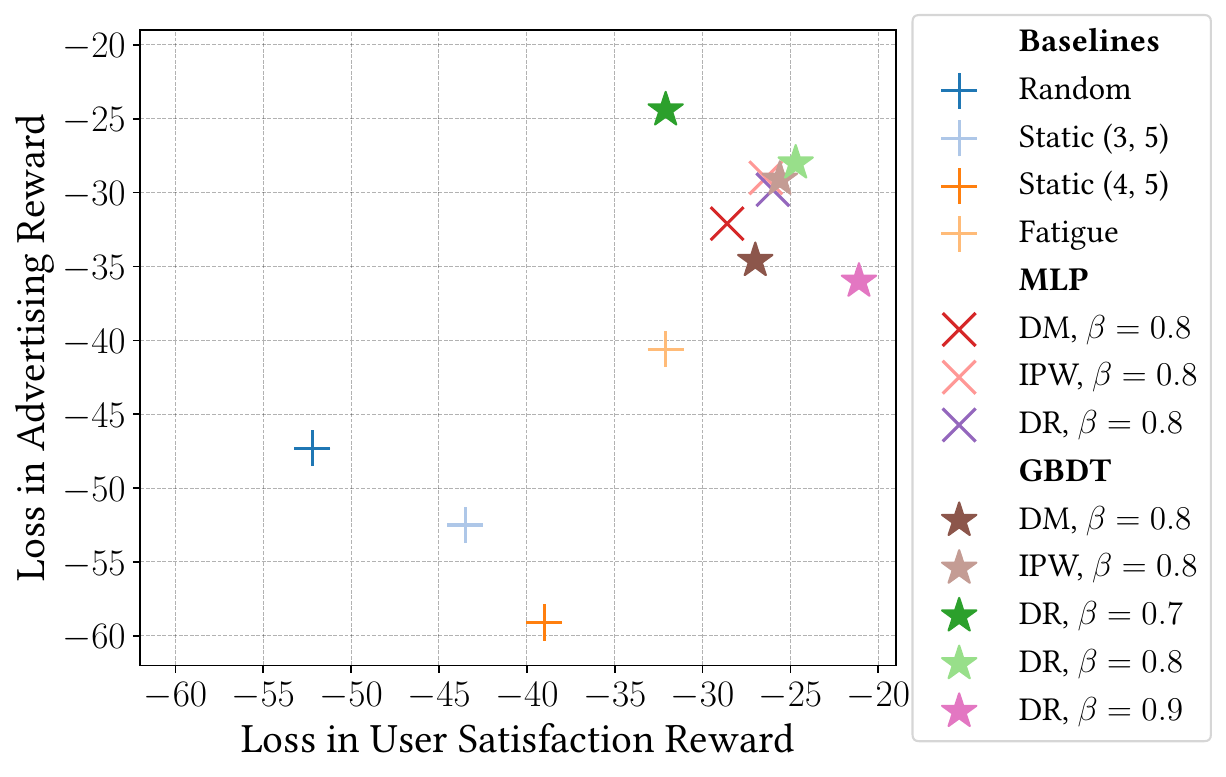}
    \caption{Offline evaluation, comparing \% loss in satisfaction and advertising objectives w.r.t. baseline policies.}
    \label{fig:offpolicy_eval}
\end{figure}

\vspace{-1em}
\subsection{Deployment \& Online A/B Experiment}
Based on the offline results, we conducted A/B tests for three policies as outlined in Table \ref{tab:ab_online_results}. Online experiments were conducted over a span of two weeks, involving approximately 5 million daily active users. At the start of the experiment, a uniform random policy was applied to 1\% of randomly selected active users in the variant, with their data logged to retrain the offline learned policies retraining.
For the remaining users, the learned policies were deployed.
To gradually transition to the learned policies for all users, the percentage of users exposed to the random policy was progressively reduced over the course of several days. The re-trained model was stored in a cloud storage bucket, assigned an incremented version, and deployment in service that loaded the updated model. The learned policy models were deployed in a Kubernetes cluster with each pod having around 64 vCPUs, 240GB of RAM and NVIDIA T4 GPUs with horizontal auto-scaling enabled based on the request rate \& cpu utilization of pod. The online p99 inference latency was approximately 25 milliseconds during peak traffic.

All the variants employed GBDT models to parameterise the learned policy.
All the variant policies used GBDTs to parameterise the learned policy, and the doubly robust estimator as their objective.
The control group was shown a personalized policy that only considers the ``fatigue score'' to increase or lower the ad-load (\textit{Fatigue} baseline in Figure~\ref{fig:offpolicy_eval}~\cite{sagtani2023quantifying}).
We tracked a number of user SAT and revenue metrics including Retention, Time spent, video plays, ads impressions, ads clicks.

variant-1, with $\beta=0.7$, exhibited negative satisfaction metrics but demonstrated a high ads objective, indicating a trade-off that favored ads and was not suitable for the platform. In contrast, both variant-2 and variant-3 showed positive SAT and ads objectives, suggesting that optimizing for multiple objectives based on context enabled us to capture user heterogeneity and achieve gains in all objectives without adversely affecting others. When comparing variant-2 and variant-3, we observed that an increase in $\beta$ value 0.8--0.9 resulted in higher user satisfaction but a lower ads objective.

The analysis of variant-2 provides valuable insights into the effectiveness of the policy. By examining the distribution of ad-slots based on ad-position and user features, we gained a deeper understanding of its performance. In terms of feed depth, we observed that for the first feed, the policy shifted nearly half of the ads to the $4^{th}$ position, while distributing the remaining ads between positions ``3'' and ``2''. This adjustment resulted in a reduced ad-load from ``3'' to ``4'' for most cases and ultimately led to an improvement in user satisfaction (SAT). For the next 3-4 feeds, the ad-load increased compensating for the revenue loss in the first feed.

Another interesting finding emerged when analyzing the policy's behavior across consecutive feeds. It exhibited an alternating pattern of high and low ad-loads. For instance, if the user had no ads in the last feed, the policy predominantly displayed first ad at $2^{nd}$ position. Conversely, if the ads count in the last feed were 2, the ads were primarily shown at positions ``5'', ``6'' or no ads at all. However, when there was only 1 ad in the last feed, the policy displayed ads of various positions, indicating a mixed approach.

Additionally, we examined the relationship between ad display and fatigue score. Notably, as the fatigue score increased, the policy reduced the frequency of ad displays and shifted towards lower ad-loads. Conversely, for users with very high fatigue scores ($ \geq 0.85$), the policy displayed higher ad-loads. Further analysis revealed that users with high fatigue scores tended to have a higher churn rate, meaning they were more likely to discontinue using the platform. As a result, the policy optimized the ads objective by increasing ad exposure to these users, as their SAT improvement was minimal.

Taking into account the insights gained from the A/B test and the analysis of variant-2, the policy was deployed in production serving 100\% traffic cross 180 million monthly active users.

\vspace{-1ex}\section{Discussion \& Future Work} \label{sec:conclusion} 
In conclusion, we have presented the ``\textit{ad-load balancing}'' problem, emphasizing the trade-off between user satisfaction and ads objectives, as well as user-level heterogeneity. Through the use of an off-policy learning framework and unbiased estimators, we have successfully learned effective policies to tackle this challenge. Our approach has resulted in improvements in both user satisfaction and ads revenue metrics for the platform. While linear scalarization helps decide the trade-off between user satisfaction and ads-objectives, we envision future extensions to research and include more sophisticated functions to have more fine grained control over these objectives.

\newpage
\section*{Ethical Considerations}
Our work aims to address the ---sometimes conflicting--- objectives that stakeholders in content marketplaces wish to optimize~\cite{Abdollahpouri2022}.
Indeed: users want to be satisfied with the platform, advertisers want successful campaigns, and the platform wants to be sustainable in the long-term.
Our method allows \emph{all} stakeholders to benefit, as evidenced by both our off- and online experiments: the learned ad-load policies move the Pareto front, improving both on user retention \emph{and} advertising revenue.

Nevertheless, as user satisfaction is notoriously hard to measure, we rely on short-term proxy signals, including those based on engagements.
It should be clear that an over-reliance on such signals can drown out others, as content likely to lead to \emph{engagement} can in fact be ``clickbait'', polarising, or otherwise harmful to users.
Even though our work merely leverages such signals as proxies to measure satisfaction under personalised ad load policies, we want to be explicit about such pitfalls.

\bibliographystyle{ACM-Reference-Format}
\bibliography{references}

\end{document}